\documentclass[usenatbib]{mn2e}

\newcommand{\be}{\begin{eqnarray}}

\newcommand{\bit}{\begin{itemize}}
\def\bkt#1{\left(#1\right)}

\def\ee{\end{eqnarray}}

\newcommand{\eit}{\end{itemize}}

\def\gtsima{$\; \buildrel > \over \sim \;$}

\def\lab{\label}

\def\ltsima{$\; \buildrel < \over \sim \;$}

\newcommand{\mmm}{\medskip}

\def\simlt{\lower.5ex\hbox{\ltsima}}
\def\simgt{\lower.5ex\hbox{\gtsima}}
\def\sub#1{_{\mbox{\scriptsize{#1}}}}

\usepackage{amsmath}
\usepackage{amsfonts}
\usepackage{amssymb}
\usepackage{graphicx}
\usepackage{epsfig}
\DeclareMathAlphabet{\mathpzc}{OT1}{pzc}{m}{it}

\title[The $M-c$ relationship of virialized halos]{The mass-concentration relationship of virialized halos and its impact on cosmological observables}
\author[] {Lindsay J. King$^{1,2}\thanks{email: ljk@ast.cam.ac.uk}$, James M.G. Mead$^{1,2}$\\ 
1. Institute of Astronomy, Madingley Road, Cambridge, CB3 0HA, United Kingdom\\
2. Kavli Institute for Cosmology, Madingley Road, Cambridge, CB3 OHA, United Kingdom.}
\date{}
\pagerange{\pageref{firstpage}--\pageref{lastpage}}
\pubyear{}

\begin{document}
\label{firstpage}
\maketitle
\begin{abstract}
A generic property of the cuspy simulated virialized halos in cold dark matter cosmogenies is that their concentration is inversely correlated with their mass. This behavior has also been confirmed in observations, although differences in the exact form and dispersion of this so-called mass-concentration relationship have been reported. Some observational studies of massive halos suggest that they are statistically over-concentrated with respect to the expectations of $\Lambda$CDM. Here we investigate the impact that various published mass-concentration relationships, both from simulations and derived from observations, would have on other cosmological observables, in particular considering upcoming surveys. We find that an integral measure of lensing shear, such as counts of peaks from halos, is very sensitive to the relationship between mass and concentration at fixed $\sigma_{8}$, and the disparity between some reported fits is much larger than the impact of uncertainty in $\sigma_{8}$ itself. We also briefly assess the impact of baryonic physics on cluster scale observables, using state-of-the-art simulations, concluding that it is unlikely to give rise to the high concentrations reported for some clusters.
\end{abstract}
\section{Introduction}
Galaxy clusters are the most massive bound structures in the universe, some with masses in excess of 10$^{15}$M$_{\odot}$.  Many observables related to clusters are sensitive cosmological probes.  
The evolution of the abundance of rich clusters  has a strong dependence on $\sigma_{8}$, the amplitude of mass fluctuations on the scale of 8$h^{-1}$Mpc (e.g. \cite{Fanbahcen}), and the covariance of cluster counts is very sensitive to primordial non-Gaussianity (\cite{ogurifnl};  \cite{cunha}). Furthermore, our understanding of the interplay between baryonic and dark matter in clusters, and of cluster formation and evolution can be improved by comparing observations of clusters with those formed in cosmological simulations that incorporate baryonic physics (e.g. \cite{mksea}). 

Numerical simulations of structure formation in cold dark matter (CDM) cosmogonies reveal that virialized dark matter halos, with masses spanning many orders of magnitude, possess a cusped universal mass density profile, independent of the initial density fluctuations and cosmological parameters \citep{nfw}. The density profile progressively steepens from the centre to the virial radius $r_{\rm vir}$, with an isothermal slope at the scale radius $r_{s}$. A measure of the mass concentration of halos is given by $c_{\rm vir}\equiv r_{\rm vir}/r_{s}$. This concentration parameter shows a good correlation with virial mass, both in simulations and in observations, and in CDM there is a generic trend that more massive halos are less concentrated. \cite{wangwhite} have demonstrated that there is also a correlation between mass and concentration in virialized halos formed in hot dark matter (HDM) cosmogenies, but with more massive halos tending to be more concentrated. Given that in CDM and HDM structure formation proceeds either via hierarchical aggregation or monolithic collapse respectively, their work indicates that mergers are not the dominant mechanism responsible for the universal profiles and distributions.                                                                                                                                                                      

Some analyses of rich galaxy clusters, in particular using gravitational lensing data, have shown concentrations higher than expected in 
$\Lambda$CDM, e.g. MS 2137-03 (e.g. \cite{gavazzi03}), Cl 0024+1654 (e.g. \cite{kneib03}) and the very well-studied Abell 1689 (e.g. \cite{broad05}; \cite{umbr08}; also see \cite{corkinclo09} for a compilation of estimates from various works). Generally speaking, factors that may contribute to this apparent excess in concentration include having a triaxial halo with the major axis close to the line of sight (\cite{oguri05}; \cite{gavazzi05}; \cite{corkin07}), where by adopting a spherical model (circularly symmetric in projection) in the analysis one can bias parameter estimates and underestimate error bars. Projection of other structures close to the line of sight can similarly result in apparently high concentration (e.g. \cite{kincor07}). Observational results are often compared with dark matter simulations, but baryons - with their more complex physics - are dominant in the centres of galaxy clusters, galaxy groups  and massive galaxies. One important process due to the presence of baryons, is that dark matter halos may undergo contraction (\cite{els}; \cite{blum86}; \cite{gnedin04}) resulting in significant deviation from the form arising in dissipationless collapse. 

Differences in the exact form of the so-called mass-concentration (hereafter $M-c$) relationship have been reported, both measured in CDM simulations and estimated from observations (e.g. \cite{bull}; \cite{comnat}; \cite{duffy08}; \cite{maccio08}; \cite{mandel08}; \cite{oguri09}; \cite{okabe09}).  For example, considering only galaxy clusters whose radial profiles are well constrained from combining weak and strong lensing data, Oguri et al. (2009) note a $7\sigma$ excess of the concentration of mass compared with expectations in $\Lambda$CDM, even including an enhancement for any bias in the samples of lensing clusters (e.g. \cite{hen07}).  \cite{mandel08} performed a statistical analysis to estimate the mass-concentration relation over three orders of magnitude in mass using the stacked weak lensing signals of galaxies, groups and clusters in SDSS. Their analysis used weak lensing data in order to avoid the central regions of halos where baryons dominate, and to minimise errors from misestimation of clusters' centres. They find a $M-c$ relationship with a slope consistent with simulations, but an amplitude that is about $2\sigma$ below theoretical expectations.

In order to make progress in understanding observed departures in the relationship between mass and concentration - beyond that expected from intrinsic scatter, and from known observational biases - in this paper we start by considering how particular $M-c$ fits derived from observations or measured in CDM simulations would impact on other cosmological observables in upcoming surveys. The Dark Energy Survey (DES\footnote{http://www.darkenergysurvey.org}), for example, is expected to contain $\sim 3,000,000$ galaxies out to $z\sim 1$  and $\sim 20,000$ clusters in excess of 2$\times 10^{14}$M$_{\odot}$, thus containing a wealth of information on virialized objects. The ingredients of the halo model formalism, used in the predictions, are summarised in Section 2, in particular the $M-c$ relationships used in the later sections. The resulting cosmic shear power spectra and the lensing selected halo counts are considered in Sections 3 and 4 respectively. In Section 5 the impact of baryons on cluster-scale mass profiles is explored, in comparison with analytic profiles of that mass and with concentrations derived from $M-c$ relationships. We summarise and discuss our findings in Section 6. Throughout we assume a flat cosmological model, and a dark energy equation of state parameter $w=-1$.

\section{The halo model description of the matter power spectrum}
In the Press \& Schechter model \citep{presch}, all matter in the universe is structured in virialized dark matter halos.
Under the halo model formalism, the halo mass function and halo clustering strength, as well as the  distribution of mass within the halos is used to estimate statistical properties of the density field (\cite{mccsil}; \cite{schber}; \cite{sel00}; \cite{peasmi}; \cite{scoea01}; \cite{cooshe}) which readily allows estimation of the impact of the properties of halos on observables (e.g. \cite{hamyosu}; \cite{neysza}; \cite{fedmos10}; \cite{pieetal}).

The matter power spectrum $P(k)$, at wavenumber $k$, is the sum of one-halo, $P^{1h}(k)$, and two-halo, $P^{2h}(k)$, terms:
\begin{equation}
P(k) = P^{1h}(k) + P^{2h}(k).
\end{equation}

The one-halo term depends on the distribution of mass in individual halos and dominates on small scales:
\begin{equation}
P^{1h}(k) = \int {\rm d}M\,\frac{{\rm d}n}{{\rm d}M} \left(\frac{M}{\bar{\rho_0}}\right)^2\,|y(k,M)|^2\,, 
\end{equation}                             
where $M$ denotes halo mass and $dn/dM$ is the halo mass function, giving the number of halos in the mass range $M\rightarrow M+dM$ at a certain redshift, per unit volume. $\bar{\rho_0}$ is the present mean mass density of the universe and $y(k,M)$ is the Fourier transform of the matter density profile of halos, normalised by their mass. 

The two-halo term arises from elements in distinct halos:
\begin{equation}                             
P^{2h}(k) =  P^{\rm lin}(k)\left[
\int {\rm d}M\,\frac{{\rm d}n}{{\rm d}M}\left(\frac{M}{\bar{\rho_0}}\right)\, |y(k,M)|b_{h}(M)\right]^{2}\,,
\end{equation}
where $P^{\rm lin}(k)$ is the linear matter power spectrum, and $b_{h}(M)$ is the halo bias that takes account of massive halos clustering more strongly than low mass halos.

\subsection{Halo mass function and bias model}

The halo mass function and bias are taken from \cite{Shethtormen99}. The mass function is given by
\begin{eqnarray}
\frac{{\rm d}n}{{\rm d}M}{\rm d}M&=&\frac{{\bar\rho_0}}{M}f(\nu){\rm d}\nu\\\nonumber
&=&\frac{\bar\rho_0}{M}A\left[1+\frac{1}{(a\nu)^{p}}\right]\left[(a\nu)^{\frac{1}{2}}\right]{\rm exp}^{-(a\nu)/2}{\rm d}\nu\\
{\rm where~~~}  \nu&\equiv& \left[\frac{\delta_c(z)}{\sigma(M,z)}\right]^2\nonumber\,,
\end{eqnarray}
where $\delta_{c}(z)$ is the cosmology-dependent critical value of the overdensity required for spherical collapse at $z$, for which we use a fitting function (see appendix of \cite{hen} and references therein)
\begin{eqnarray}
&&\delta_{c}(z)=\frac{3(12\pi)^{2/3}}{20}\left[1-0.0123\,{\rm log}(1+X^3)\right];\\\nonumber
&&X\equiv\frac{\left(\Omega^{-1}_{\rm m}-1\right)^{1/3}}{1+z}
\label{fit1}
\end{eqnarray}
appropriate for $\Omega_{K}=0$. 
$\sigma(M,z)$ denotes the variance of the linear density field, spherical top-hat smoothed in spheres enclosing mass $M$, i.e. within radius $R(M)\equiv\left(3M/4\pi{\bar\rho_0}\right)^{1/3}$. The coefficients $a=0.707$ and $p=0.3$ are fit from N-body simulations, and $A$ is a normalisation constant obtained by the requirement that all mass is in halos
\begin{equation}
\bar\rho_0=\int {\rm d}M\,M\frac{{\rm d}N}{{\rm d}M}\,.
\end{equation}
Halos are biased tracers of the background dark matter field, being associated with peaks of the initial density field. Under the peak-background split ansatz (\cite{kai84}; \cite{bbks}; \cite{colkai}; \cite{mowhi}), the large-scale halo bias can be related to the halo mass function; for the Sheth-Tormen mass function this reads
\begin{equation}
b_h(M)=1+\frac{a\nu-1}{\delta_c}+\frac{2p}{\delta_{c}\left[1+(a\nu)^{p}\right]}\,.
\end{equation}
\subsection{Halo mass density profile}
The NFW profile (\cite{nfw}) is taken as the fiducial halo matter density profile that enters as an ingredient in the halo model via its Fourier transform. The mass density $\rho$ as a function of radius $r$ is given by 
\begin{equation}
\rho(r)=\frac{\rho_{\rm s}}{(r/r_{\rm s})(1+r/r_{\rm s})^2}\,,
\end{equation}
where $\rho_s$ is the matter density at the scale radius $r_{\rm s}$. The concentration parameter $c$ is given by $c=r_{\rm vir}/r_{\rm s}$, where $r_{\rm vir}$ is the virial radius of the halo (roughly equivalent to $r_{200}$, the radius within which the mean enclosed density is 200 times the critical density of the universe at the redshift at which the halo virialized). The virial mass of a halo with the NFW profile is 
\begin{equation}
M_{\rm vir}=\frac{4\pi \rho_{s}r^{3}_{\rm vir}}{c^3}\left[{\rm log}(1+c)-\frac{c}{1+c}\right],
\end{equation}
which, from the spherical collapse model, is equivalent to
\begin{equation}
M_{\rm vir}=\frac{4}{3}\pi r^{3}_{\rm vir}\delta_{\rm vir}(z)\bar{\rho_{0}}
\end{equation}
with $\delta_{\rm vir}$ being the virial overdensity. A fitting function for the virial overdensity is (\cite{eke96}; Nakamura \& Suto 1997)
\begin{equation} 
\delta_{\rm vir}=18\pi^{2}\left(1.0 + 0.4093\,X^{2.71572}\right)\,,
\end{equation}
where $X$ is defined as in Eq.\,5. 
The mass-normalised Fourier transform of a density profile $\rho(r)$, out to the virial radius is given by
\begin{equation}
y(k,M)=\frac{1}{M}\int_0^{r_{\rm vir}}{\rm d}r \rho(r)4\pi\,r^{2}\frac{{\rm sin}(kr)}{kr}
\end{equation}
and for the truncated NFW profile this reads (e.g. \cite{scoea01}):
\begin{equation}
\begin{split}
y(k,M)=4\pi\rho_0r^{3}_{s}\Big[{\rm cos}(kr_{\rm s})\left[{\rm Ci}(kr_{\rm s}(1+c))-{\rm Ci}(kr_{\rm s})\right]\\
+{\rm sin}(kr_{\rm s})\left[{\rm Si}(kr_{\rm s}(1+c))-{\rm Si}(kr_{\mathrm s})\right]-\frac{{\rm sin}(kr_{\rm s}c)}{kr_{\rm s}(1+c)}\Big]
\end{split}
\end{equation}
where Ci and Si are cosine and sine integrals respectively
\begin{equation}
{\rm Si}(z)\equiv \int_{0}^{z}\frac{{\rm sin}(t)}{t}{\rm d}t\,;~~~~~~{\rm Ci}(z)\equiv -\int_{z}^{\infty}\frac{{\rm cos}(t)}{t}{\rm d}t\,.
\end{equation}
 For simplicity and ease of comparison with other work, we focus on the NFW profile here, although higher resolution simulations of halo formation show a softening of the profile with a logarithmic slope that decreases with radius, the so-called Einasto profile \citep{stadel}. In Section 5 deviations from the NFW profile due to the inclusion of baryons are considered.
 
The lensing properties of the non-truncated and truncated NFW halo are obtained by integrating the density profile along the line of sight to obtain the surface mass density, $\Sigma$, often given in terms of a scaled projected distance from the centre $x\equiv r'/r_{s}$ where $r'$ is the projected distance. This yields the convergence, $\kappa(x)$, when scaled by the critical surface mass density $\Sigma_{\rm crit}$ that depends on the lens and source distances. Since the surface mass density is circularly symmetric, the shear 
$\gamma$ can be obtained using the relationship $\gamma(x)=\bar\kappa(x)-\kappa(x)$, where $\bar\kappa(x)$ is the mean convergence inside $x$. See for example Bartelmann (1996) and Wright \& Brainerd (2000) for the non-truncated and Takada \& Jain (2003) for the truncated case (where the halo is assumed to extend to e.g. the virial radius). Of importance here is that $\kappa$ and $\gamma$ depend on the shape of the profile as well as the mass. For example
\cite{hamtakyos} note that $\kappa$ scales approximately $\propto c$ for $c\lesssim 5$, and $\propto c^{1.5}$ for larger $c$.

\subsection{Mass-concentration relationships}
\begin{figure}
\hspace{-0.5cm}
\epsfig{file=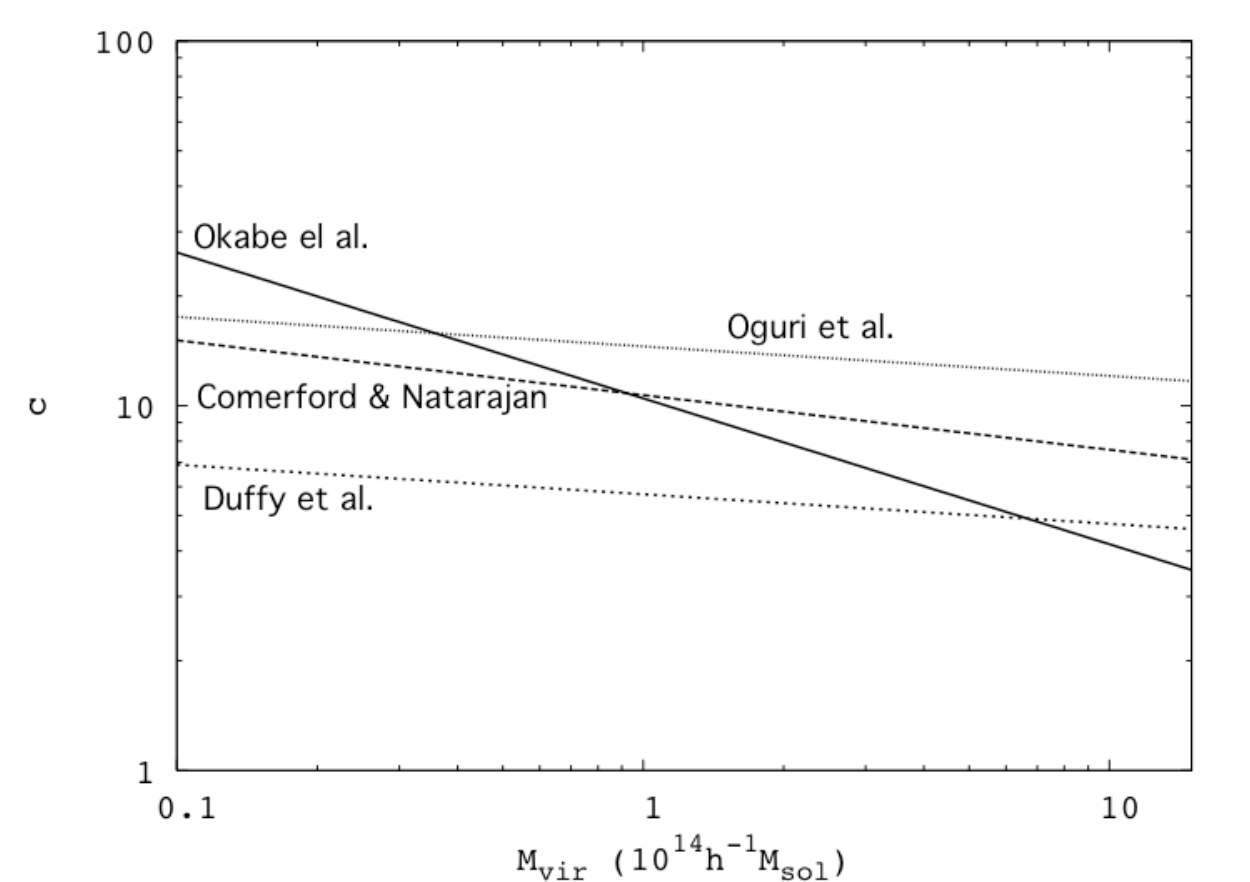, width= 9cm, angle = 0}
\caption{$M-c$  relationships  at $z=0$ as described in the text.}
\label{mcs}
\end{figure}

The relationship between the mass and the concentration of virialized halos has been determined by various authors using N-body simulations or observational data. Here several representative fits from the literature are adopted, working in terms of virial concentrations and masses as noted above. Where necessary, normalisations are recast relative to a halo of $10^{14}h^{-1}{\rm M}_{\odot}$ for ease of comparison. In Fig.\,\ref{mcs} the $z=0$ $M_{vir}-c_{vir}$ fits are collated. 
\begin{itemize}
\item{Using N-body simulations carried out assuming the WMAP 5-year cosmological parameters \citep{wmap5}, Duffy et al. (2008) derived the mean concentration of their full sample of halos (including relaxed and unrelaxed halos) as a function of mass and redshift:
\begin{equation}
c^{{\rm\small DUFFY}}(M,z) =\frac{5.72}{(1+z)^{0.71}}\left[\frac{M_{\rm vir}}{10^{14}h^{-1}{\rm M}_{\odot}}\right]^{-0.081}\,,
\end{equation} 
where $\Omega_{\rm m}=0.26$, $\Omega_{\Lambda}=0.74$, $h=0.72$ and $\sigma_{8}=0.8$. All their fits used a pivot mass of $M_{\rm pivot}=2\times 10^{12}h^{-1}M_{\odot}$ in order to minimise the covariance between the normalisation and the exponent of the mass dependence. The $1\sigma$ confidence intervals on normalisation, exponent of mass dependence and exponent of redshift dependence are roughly 2\%, 7\% and  6\% of their fit values.}

\item{
Comerford \& Natarajan (2007) presented an observed  $M-c$  relation, drawing values from the literature (both lensing and X-ray determinations for mass and concentration), as well as using newly determined masses and concentrations for 10 strong lensing clusters. Where necessary they converted their compiled values to a flat $\Lambda$CDM cosmology with $\Omega_{\rm m}=0.3$, $\Omega_{\Lambda}=0.7$ with $h=0.7$. Using this extensive sample of 62 clusters they fit a concentration normalisation, and an exponent for the mass dependence, keeping an inverse scaling with $(1+z)$ for their fit and using a pivot mass of $M_{\rm pivot}=1.3\times 10^{13}h^{-1}M_{\odot}$:
\begin{equation}
c^{{\rm\small COM}}(M,z) =\frac{10.7}{1+z}\left[\frac{M_{\rm vir}}{10^{14}h^{-1}{\rm M}_{\odot}}\right]^{-0.15}\,,
\end{equation} 
with an error of $\approx 40\%$ on the normalisation and $\approx 85\%$ on the exponent of the mass dependence. 
}

\item
{Oguri et al. (2009) considered an observed sample of 10 lensing clusters whose profiles are well-fit by a combined weak and strong lensing analysis. Assuming the WMAP 5-year cosmological parameters as above, and taking the mass and redshift scaling from the fit to N-body simulations of Duffy et al. (2008):
\begin{equation}
c^{{\rm\small OGURI}}(M,z) =\frac{14.55}{(1+z)^{0.71}}\left[\frac{M_{\rm vir}}{10^{14}h^{-1}{\rm M}_{\odot}}\right]^{-0.081}\,.
\end{equation} 
They note that there is a 7$\sigma$ excess of the concentration parameter above the prediction from the $\Lambda$CDM simulations, when the ten clusters with weak and strong lensing analysis available are combined.
} 

\item{
Okabe et al. (2010) used Subaru data to carry out weak lensing analyses of 30 X-ray selected galaxy clusters from the Local Cluster Substructure Survey\footnote{ PI: Smith; http://www.sr.bham.ac.uk/locuss}. Throughout their analysis they assumed $\Omega_{\rm m}=0.27$, $\Omega_{\Lambda}=0.73$, $h=0.72$, consistent with \cite{wmap7}. The $M-c$  relationship was derived from the 19 spectroscopically confirmed clusters with observations in 2 filters, that are well-fit by the NFW profile
\begin{equation}
c^{{\rm\small OKABE}}(M,z) =8.75\left[\frac{M_{\rm vir}}{10^{14}h^{-1}{\rm M}_{\odot}}\right]^{-0.4}\,.
\end{equation} 
The authors note that none of the massive clusters in this sample show very high concentrations, as has been reported for some clusters. They also point out that the scaling of concentration parameter with mass is tentatively steeper than predicted in simulations, though there is an error of $\approx 50\%$ on the mass exponent.  For the purposes of Fig.\,\ref{mcs}, we include an inverse dependence of $c$ on $(1+z)$ and rescale to $z=0$, given the mean redshift of their clusters.
}
\end{itemize}

\section{Weak lensing convergence power spectrum}

The power spectrum of the lensing convergence is obtained from a weighted integral of the 3D matter power spectrum as described in e.g. Bartelmann \& Schneider (2001)
\begin{equation}
P_{\kappa}(\ell)=\frac{9}{4}\left(\frac{H_{0}}{c}\right)^{4}\Omega^{2}_{\rm m}\int_{0}^{\chi_{h}}{\rm d}\chi\,P_{\rm 3D}\left(\frac{\ell}{f_{K}(\chi)},\chi\right)\frac{W^{2}(\chi)}{a^{2}(\chi)}\,,
\end{equation}
where $\chi$ is the comoving distance, $\chi_{h}$ is the comoving distance to the horizon, and $f_{K}(\chi)$ is the comoving angular diameter 
distance.\footnote{Since the spatial curvature of the universe is taken to be $K=0$ it follows that $f_{K}(\chi)\equiv\chi$.} The scale factor of the universe $a(\chi)$ is normalised to unity today. The wave vector $k$ is related to the angular wave vector $\ell$ and $f_{K}(\chi)$ through $k=\ell/f_{K}(\chi)$. The 3D matter power spectrum is calculated using the halo model formalism as described in Section 2.
$W(\chi)$ is the weighting function that takes into account the redshift distribution of the sources from which the weak lensing signal is being measured, along with the relative geometry of these sources and the density fluctuations responsible for lensing:
\begin{equation}
W(\chi)=\int_{\chi}^{\chi_{h}}{\rm d}\chi'{\cal G}(\chi')\frac{f_{K}(\chi'-\chi)}{f_{K}(\chi')}\,,
\end{equation}
where ${\cal G}(\chi')$ is the normalised source distance distribution. Note that the lower integration limit arises since only density fluctuations at $\chi < \chi'$ contribute to the convergence.  

\begin{figure}
\hspace{-1cm}
\epsfig{file=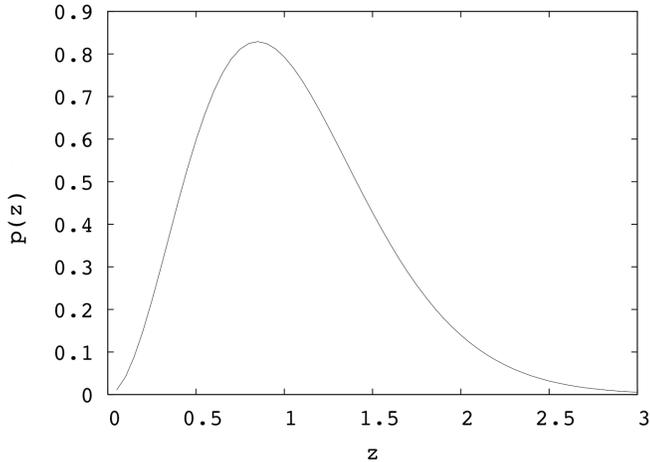, width= 10cm, angle = 0}
\caption{The fiducial redshift distribution corresponding to Eq.\,\ref{zdist} with ${\bar z\approx 1.05}$.}
\label{figzdist}
\end{figure}

We consider sources distributed in redshift according to a normalised probability distribution (Brainerd et al. 1996)
\begin{equation}
p(z)=\frac{\beta z^{2}}{\Gamma(3/\beta)z^{3}_{0}}{\rm exp}\left[-(z/z_0)^\beta\right]\,,
\label{zdist}
\end{equation}
where $z_0$ is related to the depth of the observations and $\beta$ controls the fall-off at high redshift.
For example, as in Lombardi, Schneider \& Morales-Merino (2002) when $\beta=1.5$ and $z_0=0.7$, the distribution has ${\bar z}\approx1.05$; this is taken to be the fiducial distribution and is shown in Fig.\,\ref{figzdist}. The mean redshift is representative of current cosmic shear surveys.

\begin{figure}
\hspace{-1cm}
\epsfig{file=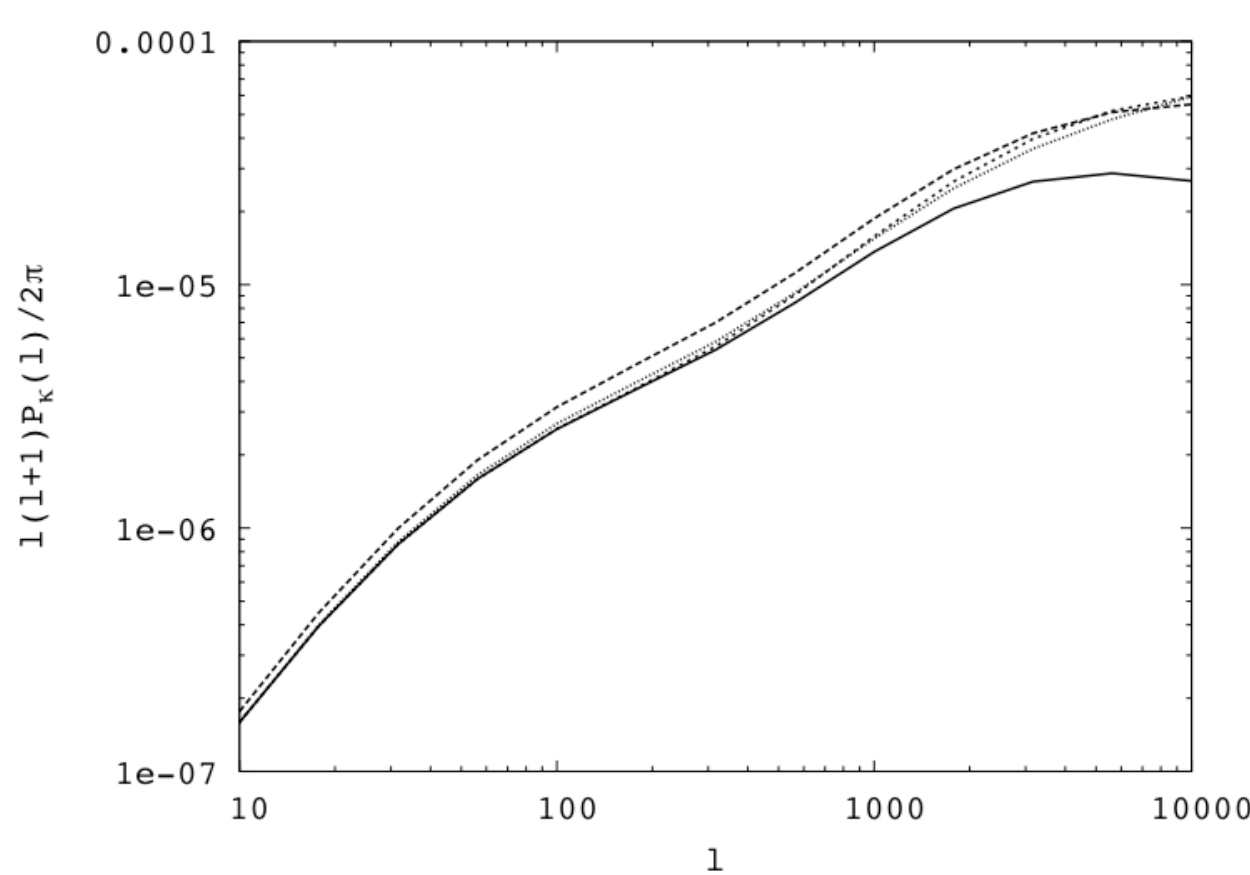, width= 10cm, angle = 0}
\caption{The dimensionless convergence power spectra for the Duffy et al.  (lower solid line), Comerford \& Natarajan (upper heavy dashed line),  Oguri et al. ( shorter dashed line) and Okabe et al. (dotted line) $M-c$ relationships with their respective cosmologies (and using the same $\sigma_{8}=0.8$ for the three observational relationships).}
\label{powspecs}
\end{figure}

In Fig.\,\ref{powspecs} the cosmic shear power spectrum for each of the prescriptions for the $M-c$  relationships is plotted, assuming the fiducial redshift distribution. 
The uncertainty in the power spectrum, assuming Gaussian errors, is given by \cite{kaiser}
\begin{align} \Delta P_\kappa(\ell)=\sqrt{2\over (2\ell+1) f\sub{sky}}\bkt{P_\kappa(\ell)+{\langle{\gamma\sub{int}^2\rangle}\over \bar{n}}}\lab{intrin},
\end{align}
where $f_{sky}$ is the fraction of the sky covered by the survey, the mean-square intrinsic ellipticity is $\langle{\gamma\sub{int}^2\rangle}$ and the number of galaxies per steradian for which an ellipticity can be measured is 
$\bar{n}$. The first term is the cosmic variance, dominant on large scales, and the second term is Poisson noise that dominates on small scales. We take $\langle{\gamma\sub{int}^2\rangle}^{1/2}=0.4$. 

\begin{figure}
\epsfig{file=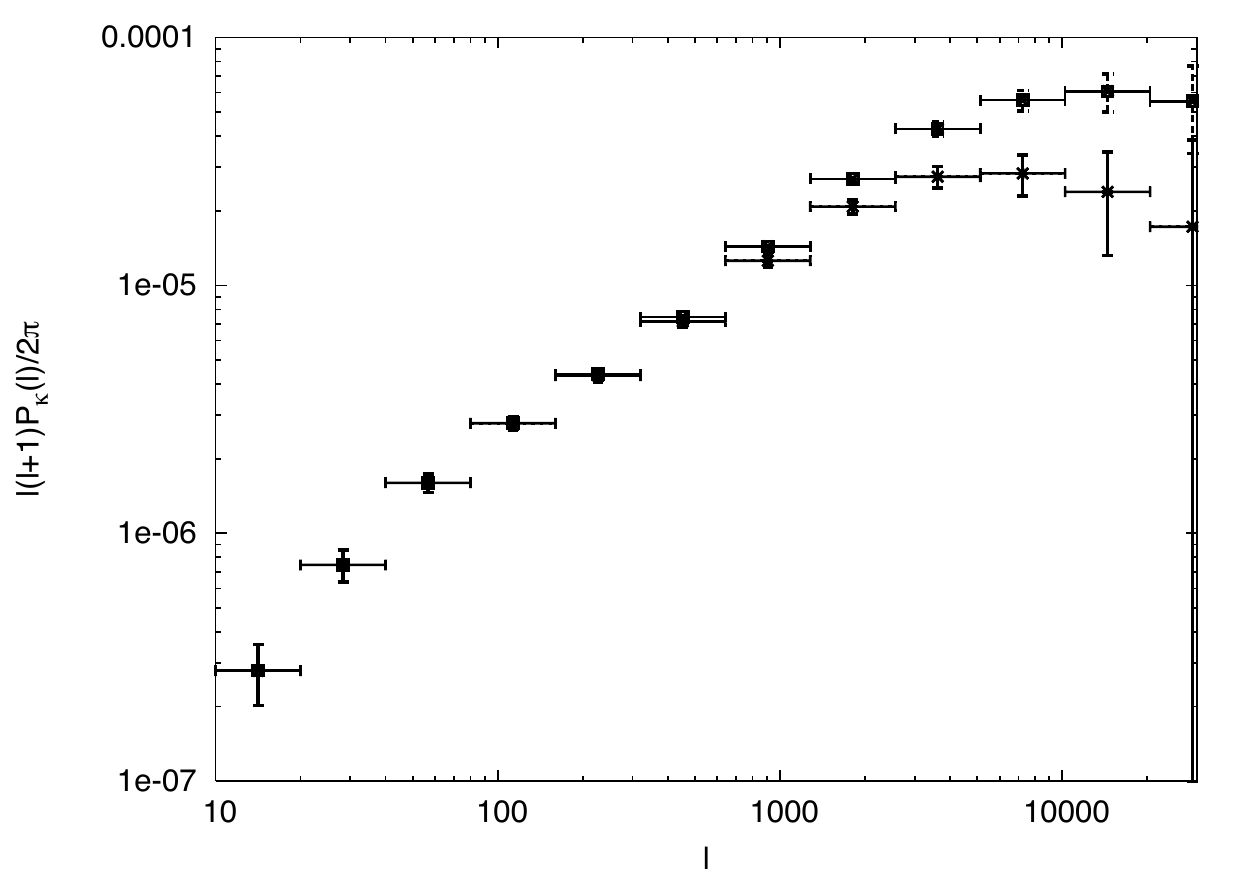, width= 9cm, angle = 0}
\epsfig{file=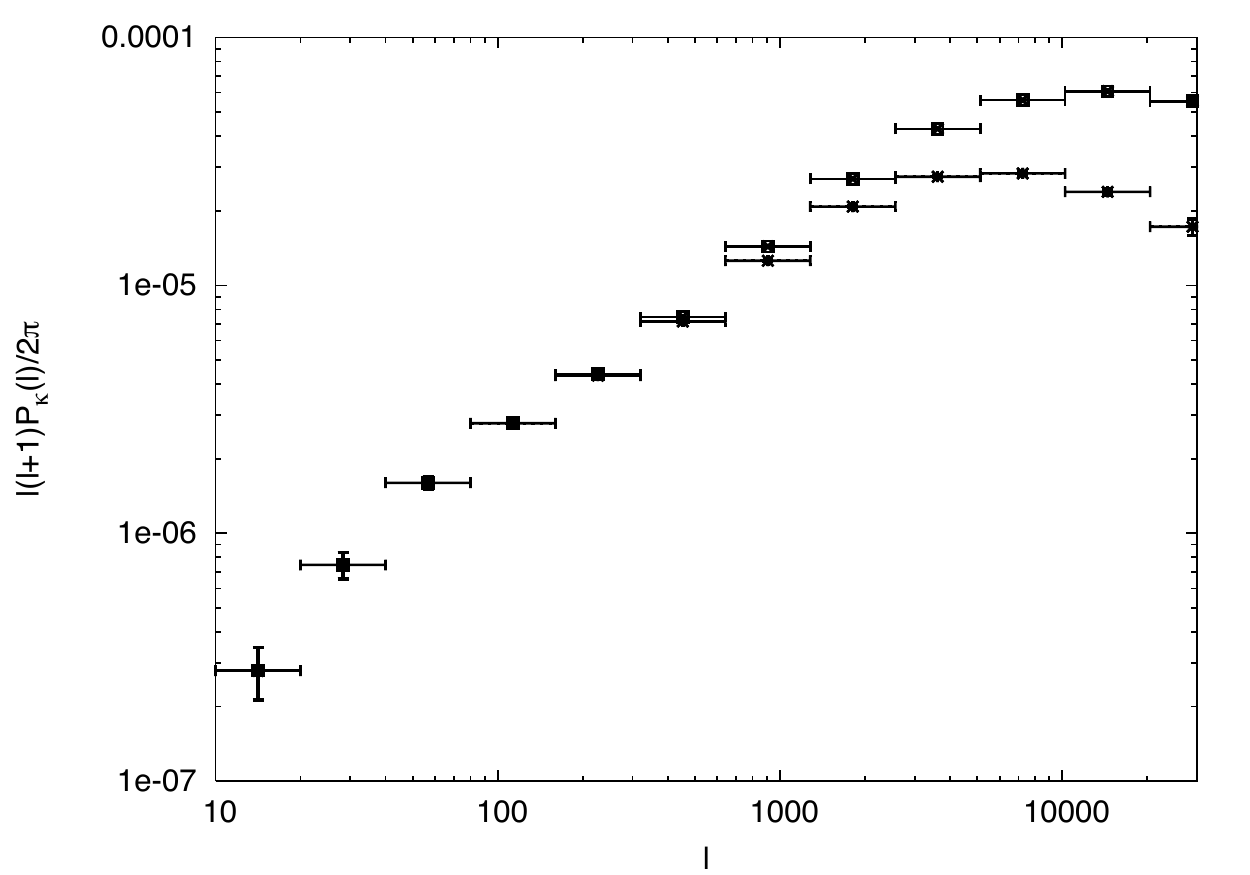, width= 9cm, angle = 0}
\caption{The upper and lower panels show the band power dimensionless convergence power spectra for the Duffy et al. (lower set of points in both panels) and Oguri et al. (upper set of points in both panels) $M-c$  relationships. The upper panel is for a source number density of 10 arcmin$^{-2}$ and sky coverage of 5000 sq. degrees. The lower panel is for a source number density of 165 arcmin$^{-2}$ and the same sky coverage. The horizontal bars denote the binning (described in the text) used to derive the errors denoted by the vertical bars.}
\label{powspecsdiffs}
\end{figure}

We now focus on the Duffy et al. (from simulations) and Oguri et al. (from observations) power spectra, since they assume the same cosmology in their analysis, and are representative of relationships reported in the literature. Fig.\,\ref{powspecsdiffs} reproduces the power spectra from Fig.\,\ref{powspecs}, now with error bars corresponding to a survey of 5000 sq. degrees and (i) a number density of 10 arcmin$^{-2}$, characteristic of the upcoming DES, and (ii) to a futuristic survey with the same area but a number density of 165 arcmin$^{-2}$ representative of deep space-based observations (e.g. the Hubble Deep Field as used in \cite{huterer}). In practice, band power estimates of the power spectrum would be made, for which we have taken 13 equal logarithmic width bins between $l=10$ and $8\times 10^{4}$ as in \cite{schn02}. The error estimate of Eq.\,\ref{intrin} correspondingly scales as $\sqrt{1/\Delta l}$. As noted above, this expression assumes Gaussian errors, whereas there is also a non-Gaussian contribution arising from mode-coupling due to non-linear clustering that becomes important at $l\sim 1000$ for a survey of the depth we consider (e.g. \cite{scoea99}), and from the statistical fluctuations in the number of halos of a certain mass which are sampled in a finite survey volume (\cite{hukra}). As noted by \cite{scoea99}, the line-of-sight projection relevant for weak lensing tends to reduce the importance of non-Gaussianity compared with what is expected in 3-D. \cite{takjai} estimate a factor of $\sim2$ degradation in the cumulative signal-to-noise ratio of the power spectrum amplitude.  We return to this point in Section 6 when the results are discussed.

\section{Halo counts}
\cite{guido} have shown that the weak lensing selected number counts of halos is a sensitive probe of cosmology, and this was tested by \cite{reblin} using ray-tracing through N-body simulations. Other authors have further explored various aspects of cluster detection using weak lensing (e.g. Hamana et al. (2004); \cite{diehar}; \cite{mar10}; see also the overview in the Introduction of \cite{mar10}), and searched for mass concentrations in data (e.g. \cite{witt06}; \cite{gava07}; \cite{mischa07}). 

We now estimate the number counts due to the most massive halos, adopting the Duffy et al. and Oguri et al. $M-c$  relationships. Again we assume that halos can be described by NFW profiles, truncated at their virial radii. Selecting halos by integral measures of their gravitational lensing shear is sensitive not only to their mass function, but to how their mass is distributed, so one expects a difference in halo counts that depends on how concentrated a halo is at fixed mass. We consider a Gaussian smoothed convergence map, that results in halos with peak convergence $\kappa_{s}$ after smoothing with a filter of scale $\theta_{s}$. We ask how many peaks would have a signal-to-noise $\nu$ above a threshold $\nu_{\rm th}$, where $\nu$ is defined as $\kappa_{s}/ \sigma_{s}$, and $\sigma_{s}$ is the noise inside the same filter arising from having a finite number density of galaxies (${\bar n}$) from which to measure the lensing signal, and from their intrinsic ellipticities (Kaiser \& Squires 1993)
\begin{equation}
\sigma_{s}^2=\frac{\gamma_{{\rm int}}^2}{4\pi \bar{n}\theta_{s}^2}\,.
\end{equation}

For sources at redshift $z$, halos with and an assumed  $M-c$  relationship, smoothing on a scale of $\theta_{s}$ and a threshold 
of $\nu_{\rm th}$ 
\begin{equation}
N(\nu>\nu_{\rm th})=\frac{1}{4\pi}\int {\rm d}\chi~\frac{{\rm d}V}{{\rm d}\chi} \int dM~\frac{dn}{dM} {\cal{H}}(\nu(M,c,z)-\nu_{\rm th}),
\end{equation}
where ${\rm d}V/{\rm d}\chi$ is the comoving volume element, ${\cal{H}}(x)$ is the Heaviside step function (${\cal{H}}(\nu(M,c,z) -\nu_{\rm th})=1$ iff $x>0$) through which the dependence on the $M-c$  relation enters.

\begin{figure}
\epsfig{file=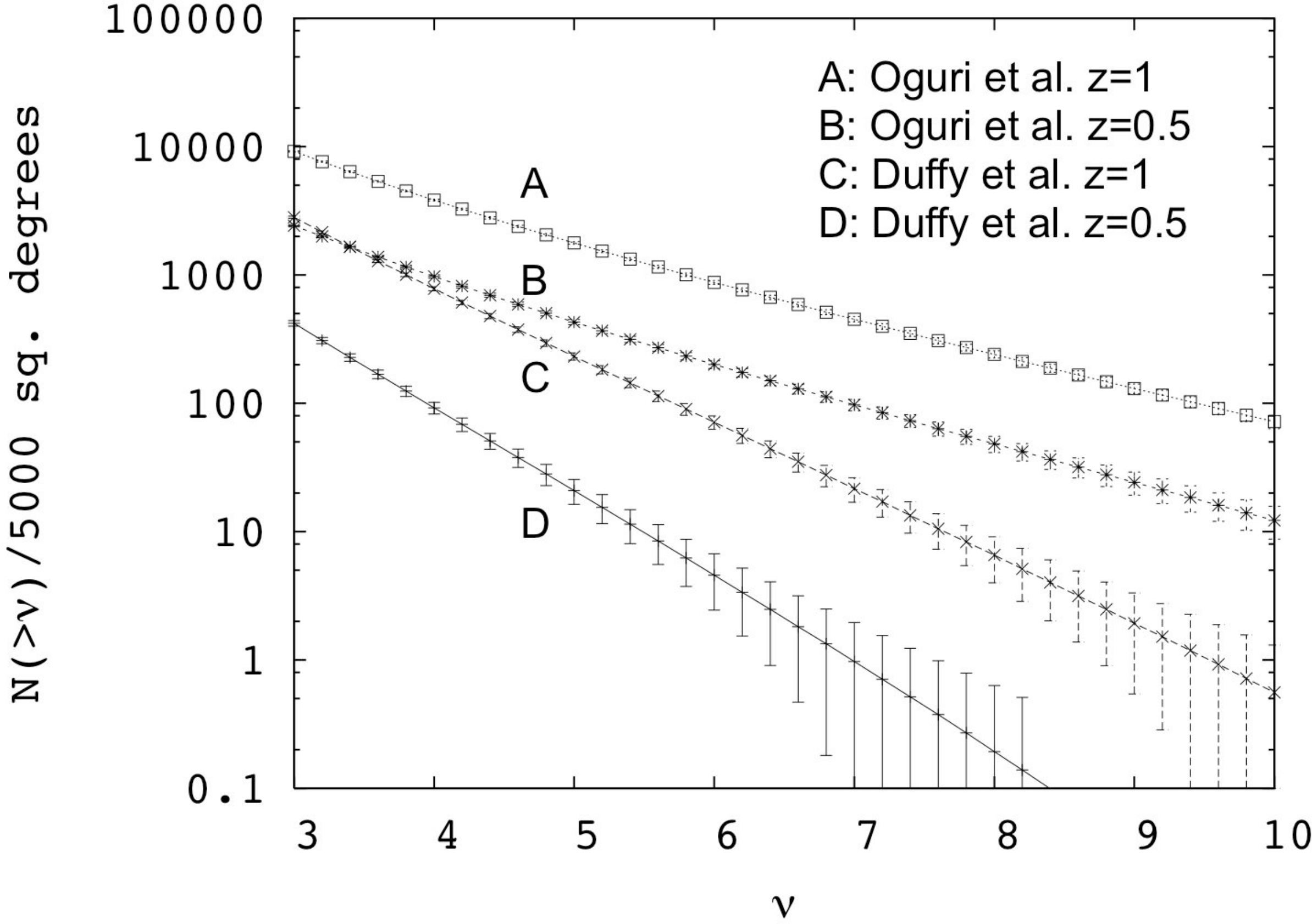, width= 8cm, angle = 0}
\caption{The number of peaks detected above a threshold $S/N=\nu$, for the Duffy et al. and Oguri et al. $M-c$  relationships for a source number density of 10 arcmin$^{-2}$ and sky coverage of 5000 sq. degrees, assuming sources at $z=1$ and $z=0.5$; the key shows the $M-c$  relationship and source redshift, with the middle line of each set corresponding to the estimated number of detections  and Poisson errors indicated by the error bars.}
\label{counts2}
\end{figure}
We consider a survey of 5000 sq. degrees and a number density of 10 galaxies per arcmin$^{2}$ usable for shear measurements, 
with the data Gaussian smoothed on a scale of 1$\arcmin$, appropriate to the scale of massive halos. In practice a filter that is better matched to the profile expected could also be employed as in e.g. \cite{bks}, or as discussed in Section 6.2 an optimal filter tailored to minimise confusion of real halos with large scale structure, e.g. \cite{mat05}, but for the purposes of illustrating the differences between $M-c$ relationships, Gaussian smoothing suffices. We first assume that sources are at each of $z=0.5$ and $z=1.0$ and fix the cosmology at the WMAP 5-year parameters ($\sigma_{8}=0.8$) that were used in the Duffy et al. simulations, and adopted in the Oguri et al. analysis. The results are shown in Fig.\,\ref{counts2}. 

\begin{figure}
\epsfig{file=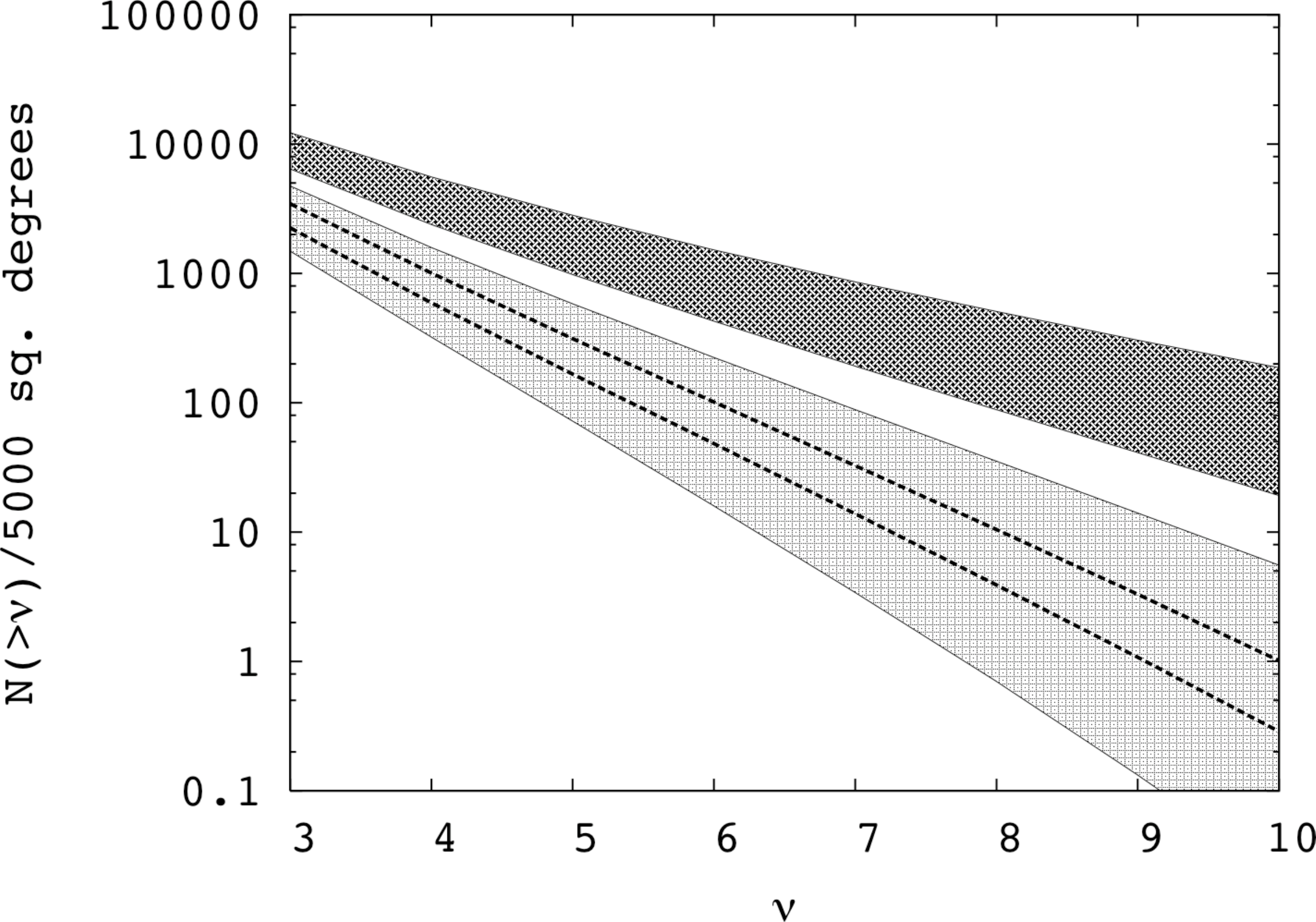, width= 8cm, angle = 0}
\caption{The number of peaks detected above a threshold $S/N=\nu$, for the Duffy et al. and Oguri et al. $M-c$  relationships for a source number density of 10 arcmin$^{-2}$ and sky coverage of 5000 sq. degrees, assuming sources at $z=1$. For the WMAP 5-year parameters ($\sigma_{8}=0.8$) the lower (upper) shaded regions indicate $\pm$1$\sigma$ offsets in concentration normalisation (from intrinsic scatter) in the Duffy et al. (Oguri et al.) $M-c$ relationships. For the Duffy et al. relation, the dotted lines indicate the impact of offsets in $\sigma_{8}$ approximately corresponding to the 2$\sigma$ bounds ($\sigma_{8}=0.77, 0.83$) from Komatsu et al. (2010).}
\label{counts3}
\end{figure}

Assuming a scatter in the $M-c$  relations of $\sigma(\log_{10}c)=0.15$ as estimated in Duffy et al. (2008), Fig.\,\ref{counts3} shows predictions for $\pm1\sigma$ offsets in the mean relations, now focusing on sources at $z=1$. Finally, the comparative sensitivity to $\sigma_{8}$ is also illustrated in Fig.\,\ref{counts3} with predictions for the Duffy et al. $M-c$ relationship, and now taking $\sigma_{8}=0.77, 0.83$, an offset roughly consistent with the current $2\sigma$ bounds from \cite{wmap7}.

\begin{figure}
\epsfig{file=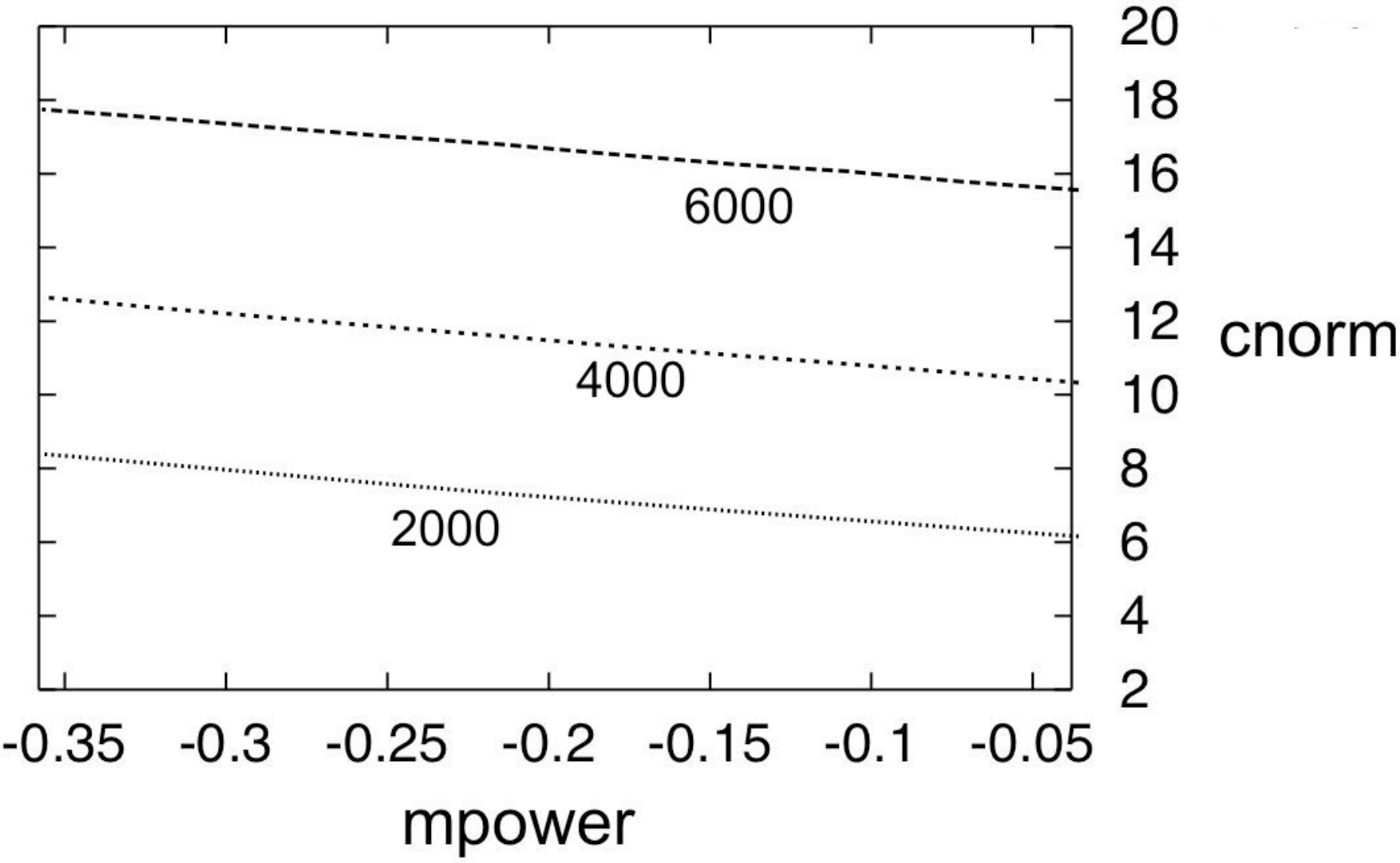, width= 8cm, angle = 0}
\\\\
\epsfig{file=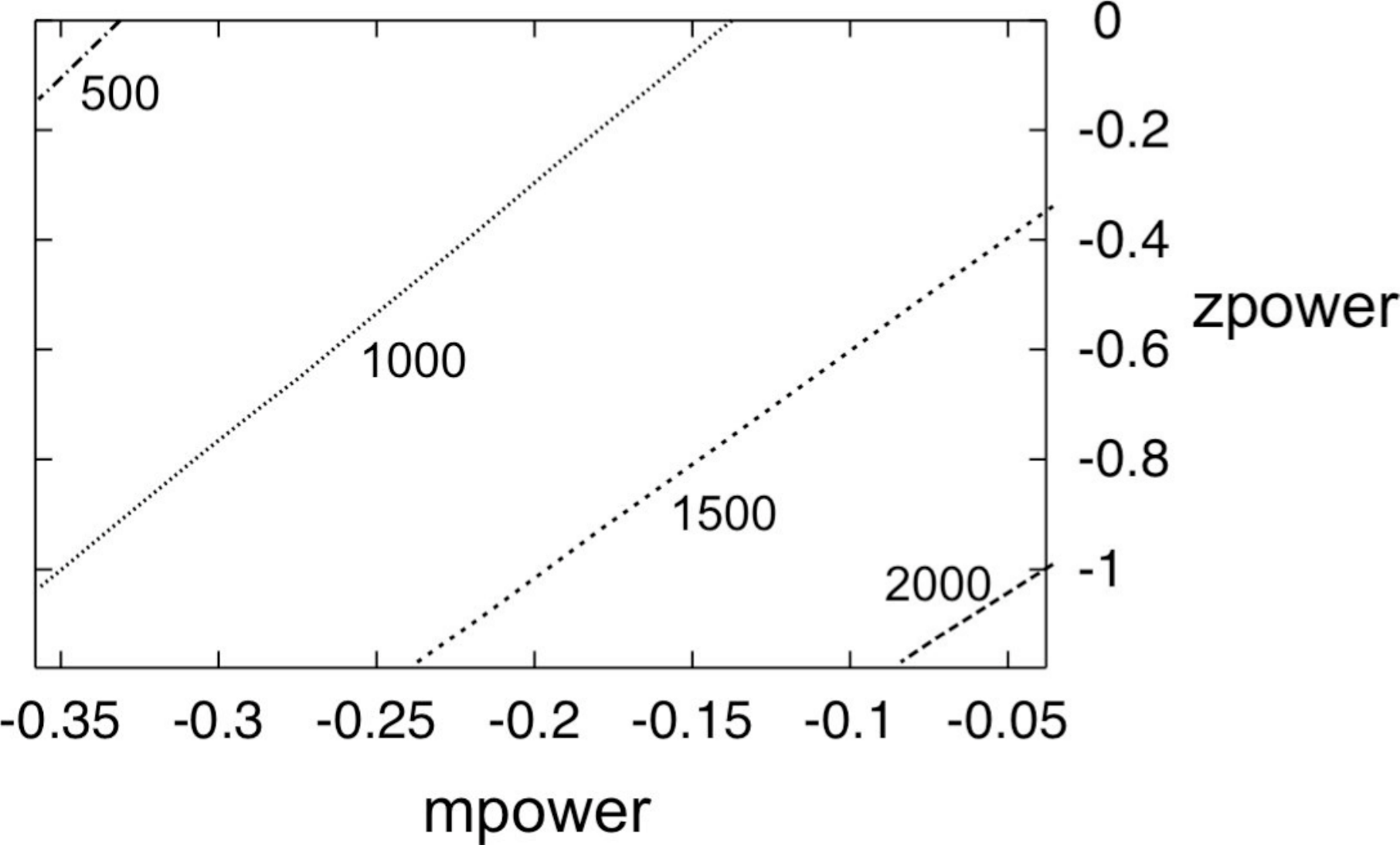, width= 8cm, angle = 0}
\\\\
\epsfig{file=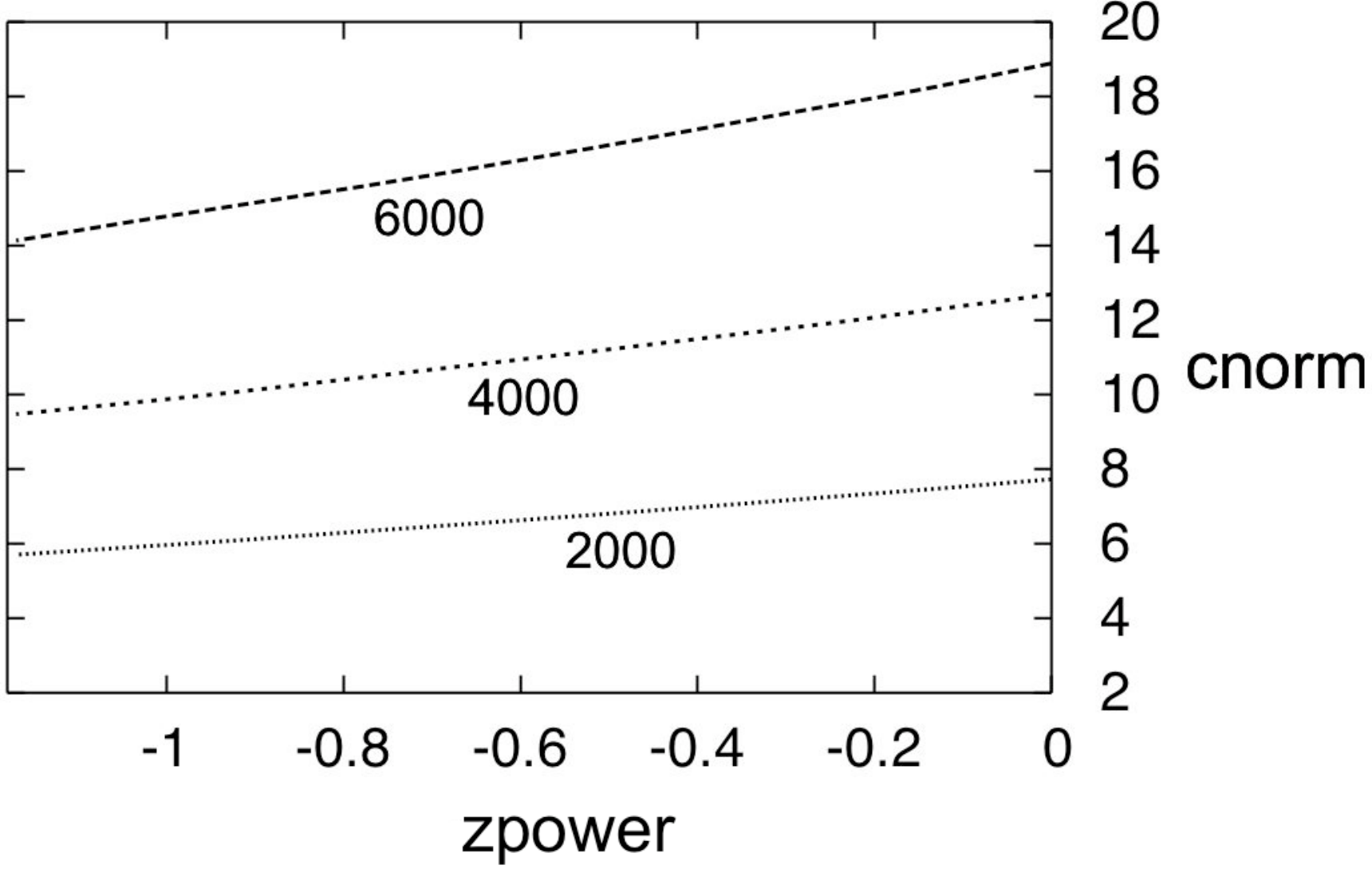, width= 8cm, angle = 0}
\caption{The number of peaks detected above a threshold $S/N=\nu_{\rm th}=4$ for a general $M-c$  relationship described in the text, for a source number density of 10 arcmin$^{-2}$ and sky coverage of 5000 sq. degrees, assuming sources at $z=1$. The upper, middle and lower panels fix zpower=-0.71, cnorm=5.72 and mpower=-0.081 respectively, and the labels on the contours denote the number of peaks.}
\label{counts4}
\end{figure}

We keep sources at $z=1$, and for $\sigma_{8}=0.8$, $\Omega_{\rm m}=0.26$, consider the dependence of counts on various of the fit parameters, for a general $M-c$ relationship 
\begin{equation}
c(M,z)={\rm cnorm}\,(1+z)^{\rm zpower}\left(\frac{M}{10^{14}h^{-1}M_{\odot}}\right)^{\rm mpower}\,.
\end{equation}
Note that this is for the purposes of illustrating how the counts would vary for changes in two of the parameters, and that some of the combinations are rather extreme. For a threshold $\nu_{\rm th}=4$ the counts are shown in Fig.\,\ref{counts4}.

\section{Baryonic physics}
The presence of baryons will cause differences in the distributions of concentrations measured from dark matter only simulations, and derived from observational data. One important process is baryonic cooling that can lead to adiabatic contraction of dark matter halos, significantly modifying their mass distributions (\cite{els}; \cite{blum86}; \cite{gnedin04}; \cite{white04}). 

Baryonic physics is likely to be most important for lower mass objects ($\lesssim 10^{14}M_{\odot}$), due to their shorter cooling timescale (\cite{silk77}; \cite{reeost77}; \cite{whiree78}). A stark illustration of this is provided by the failure to reproduce the image separation distribution of galaxy-scale gravitational lens systems, when the mass function is estimated using Press-Schechter theory (or its extensions), normalised to match the local abundance of massive clusters (see \cite{kocwhi01} and references therein). 
Based on gravitational lensing and stellar dynamics constraints from the Sloan Lens ACS Survey (SLACS; e.g. \cite{bolton06}) massive ($\sim$L$_{*}$) early-type galaxies - that tend to be central galaxies in a group or cluster, or lack more luminous neighbours - on average have isothermal  density profiles on scales of a few kpc, with a small scatter of $\lesssim 10\%$ (e.g. \cite{koop09}. This is well inside the scale radius of a typical object in their sample, where in the absence of baryons the profile is expected to be flatter than isothermal. Satellite galaxies - with more luminous companion(s) - tend to have marginally steeper mass-density profiles (e.g. \cite{auger08}; \cite{treu09}), which from N-body simulations is consistent with tidal truncation by an external potential (\cite{dobkinfel07}). \cite{schulz} fit NFW profiles to stacked weak lensing measurements in the outskirts of SDSS early-type galaxies, and find that the mass inside the half light radius directly obtained from stellar velocity dispersion observations far exceeds what is expected from the extrapolation of the weak lensing-derived profile, even when the contribution of stars is also accounted for. This is consistent with modification of dark matter profiles by baryons. 
 
Focusing on cluster scales, since these objects contribute to the halo counts discussed in Section 4, we now illustrate the differences between having dark matter only and baryonic physics (with cooling and star formation, and with and without AGN feedback). We use the total mass profiles of a $\sim 3\times 10^{14}\,{\rm M}_{\odot}$ galaxy cluster from the simulations of \cite{puchwein08} simulated with different types of physics, using the numerical prescription described in \cite{deb07}. The stellar mass fraction of the run with AGN feedback is consistent with observations (\cite{puchwein10}). Without AGN feedback the stellar mass fraction in the cluster's centre is too high, and the contraction of the dark matter is greater than seen in the simulations with feedback.  Other authors have also recently noted that when feedback is included at a level that yields stellar mass fractions consistent with observations, the central density is less enhanced than previously thought. This has been confirmed on group scales by \cite{duffy10}, where in fact the concentration parameters in the runs with baryons can actually be less than in the dark matter runs. 

\begin{figure}
\hspace{0cm}
\epsfig{file=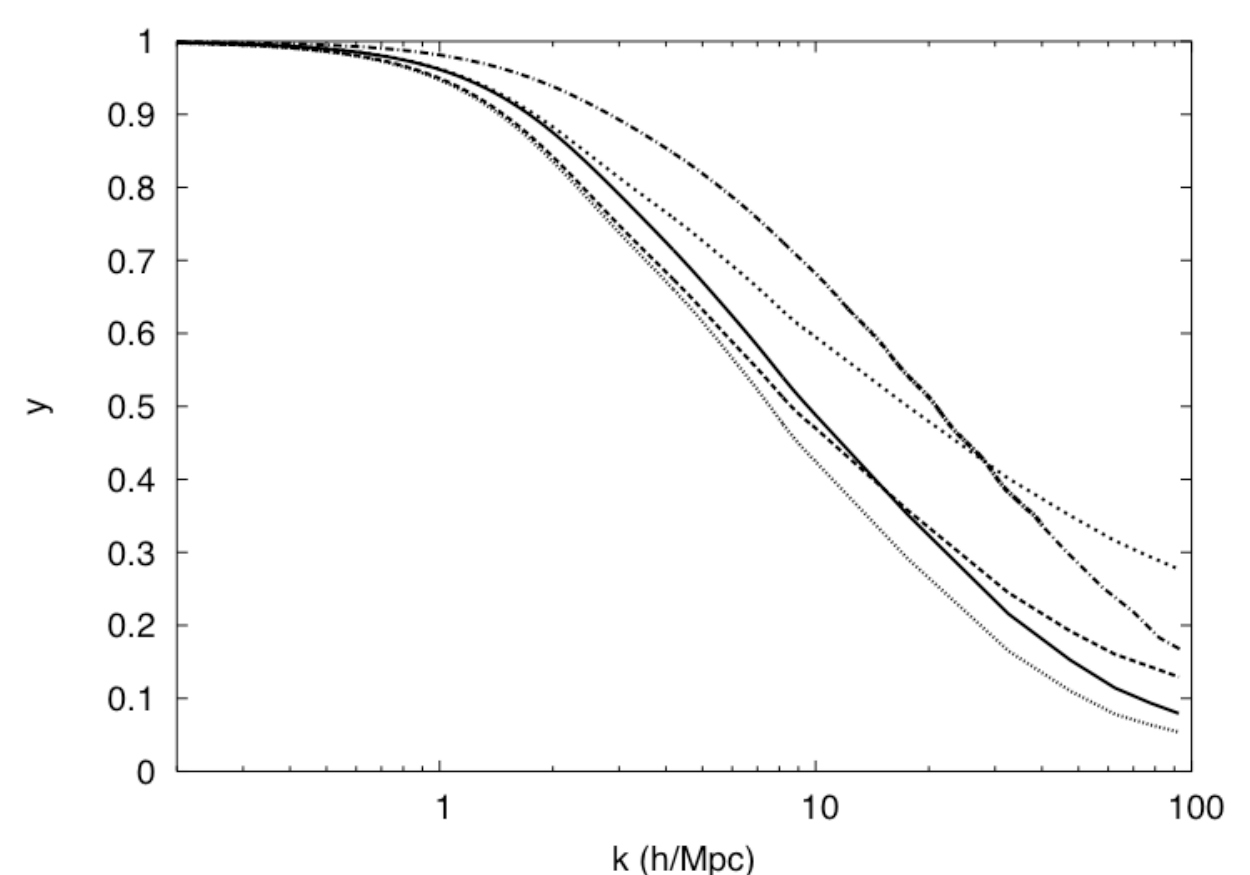, width= 9cm, angle = 0}
\caption{The mass-normalised Fourier transform of various profiles described in the text. In order of intersection with k=100\,$h$/Mpc, starting with the uppermost, the curves correspond to the simulated cluster with baryonic physics excluding AGN feedback, the analytic NFW with $c^{\rm OGURI}$, the simulated cluster with baryonic physics including AGN feedback, the analytic NFW with $c^{\rm DUFFY}$ and the simulated cluster with dark matter only.}
\label{baryoys}
\end{figure}

In Fig.\,\ref{baryoys} we plot the the mass-normalised FFT, $y$, of the spherically averaged density profile for each of the dark matter and baryonic physics runs without and with AGN feedback. We also plot the FFTs of analytic NFW profiles with the same virial radius, but with concentrations calculated using the mean for clusters of that mass from the Duffy et al. and Oguri et al. $M-c$ relationships.  

\section{Discussion and Conclusions}

Halos formed in CDM simulations have concentrations that are anti-correlated with mass. This $M-c$ relationship has been reported for both simulations and observations, with some differences in the normalisation and in the evolution with redshift. We compiled several $M-c$ relationships from the literature, and then focused on two representative fits from simulations (Duffy et al. 2008) and observations (Oguri et al. 2009) that have the same assumed cosmology. Our aim was to examine how, if objects in the universe are described by one of these forms, this would manifest in other cosmological observables measured in future surveys. The predictions have been undertaken in the framework of the halo model.

\subsection{Cosmic shear power spectrum}
In Section 3, the cosmic shear power spectra were derived for a near-future survey (with approximately the characteristics of DES) and a more futuristic space-based survey, for halos following the Duffy et al. and Oguri et al. $M-c$ relationships (see Fig.\,\ref{powspecsdiffs}). Since the assumed cosmology was the same for each of these, on large scales the signals are identical. Differences in halo properties are manifest above $l\sim 1000$ where, as noted earlier, non-Gaussian errors are likely to be important. With the typical $\sim 2$, or more pessimistic $\sim 5$ degradation in the cumulative signal-to-noise expected \citep{takjai}, the normalisation of the $M-c$ relationship would be difficult to directly constrain, until deep space-based data become available (lower panel of Fig.\,\ref{powspecsdiffs}). However, as proposed by \cite{zen}, self-calibration of cosmic shear tomographic data could be used to simultaneously extract information on halo structure as well as on cosmological parameters, in the regime where $l<3000$, until non-Gaussian errors dominate. Given that the differences in the normalisations of the two relationships considered here are apparent at $l\sim 1000$, it could well be that combining tomography with pushing to higher $l$ would allow useful constraints on the $M-c$ relationships for even the shallow survey. Zentner et al. note that the concentrations of $10^{14}h^{-1}$M$_{\odot}$ halos at $z\sim 0.2$ could be constrained to better than 10\%.

\subsection{Halo counts}
The numbers of  lensing shear-selected mass peaks detected above a certain threshold were estimated in Section 4. These turn out to be rather disparate for the two $M-c$ relations considered, already in the near-future survey as shown in Fig.\,\ref{counts2}. In Fig.\,\ref{counts3} we show the results for the peak counts assuming that the normalisations of the mean relations are offset by $\pm1\sigma$, based on the intrinsic dispersion that likely reflects formation history. For the Duffy et al. relation, we also considered changes in $\sigma_{8}$ from 0.8 to 0.77 and 0.83, at a level consistent with the 2$\sigma$ bound from \cite{wmap7}, so changing the halo mass function. The impact on number counts of this departure in $\sigma_8$ is much less than the disparity between the different $M-c$ relations. Our fiducial cosmological parameters are fixed at WMAP 5-year values (for consistency with the $M-c$ relations), with errors on $\sigma_{8}$ consistent with the current 7-year values of \cite{wmap7}; the errors on parameters will be substantially smaller after the Planck mission \footnote{see http://www.rssd.esa.int/SA/PLANCK/docs/Bluebook-ESA-SCI(2005)1\_V2.pdf}.

In this work we have employed a Gaussian filter, applied to the convergence field. In practice, one would filter the reduced shear, the direct observable from the ellipticities of weakly lensed galaxies. 
Various filters have been developed and applied to simulations and to real data. \cite{sch96} and \cite{sch98} proposed the aperture mass statistic, with a compensated weight function that gives an optimized signal-to-noise for approximately isothermal mass profiles. \cite{pad03}, \cite{mischa04a} and \cite{mischa04b} suggested filter functions that more closely follow the NFW profile. \cite{henspe} developed a tomographic matched filtering scheme for cluster detection, that in addition uses redshift information for the source galaxies. With and without tomographic information, they consider the performance of the Gaussian filter, of the aperture mass statistic - with a commonly used polynomial filter from the family in Schneider et al. (1998), of the Padmanabhan et al. (2003) filter approximately tuned to the NFW profile, and of the latter modified by an exponential cut-off to minimise the impact of large-scale structure. Using cosmological simulations to generate weak lensing data, for the case without tomographic information, \cite{henspe} find that at moderate detection significance (S/N $\sim 3.5-4.5$) the truncated NFW filter is most effective at cluster detection, followed by the Gaussian filter. For the highest  detection significance, the aperture mass statistic performs best. For a small range of detection S/N, the Gaussian filter performs best. 

Unless the S/N threshold is chosen to be high ($\sim 4-5$ depending on the specifications of the survey), or equivalently for high mass clusters, the contamination from other structure close to the line of sight leads to the detection of significant spurious peaks in weak lensing data (e.g. \cite{rebbar}; \cite{mewhlo}; \cite{whvama}; \cite{hamtakyos}; \cite{henspe}). Optimal filtering for the detection of clusters, as proposed by \cite{mat05}, dramatically reduces the contribution from spurious peaks arising from the large scale structure and from other noise due to the intrinsic ellipticity dispersion and positions of galaxies. More recently, \cite{mat10} developed an analtyic method suitable for quantifying the degree to which spurious detections are present in a weak lensing survey, for various types of filters. Crucially, by using N-body simulations, they also confirm that using optimal filters renders the contribution from spurious peaks due to large scale structure low on all filter scales, increasing the number of real detections by up to an order of magnitude. For the analysis of data from a particular survey, an estimate of spurious peaks could be made using the formalism of \cite{mat10} (Table A.1), or by the more time consuming optimal filtering of synthetic weak lensing data sets derived from ray-tracing through cosmological N-body simulations, with appropriate observational factors for the survey such as the number density of galaxies and their redshift distribution from which the shear signal can be estimated.
What one should also keep in mind is that, as discussed by \cite{hamtakyos} and \cite{fanshaliu}, peaks arising due to noise are also more likely to appear in the vicinity of real clusters. Further, in the context of cosmological parameter constraint, which is not the subject of this paper, it has been shown by \cite{diehar} and \cite{wahama} that the number and properties of {\em all} the peaks detected in a weak lensing survey, including spurious peaks due to large-scale structure, also contain important information.

\begin{figure}
\epsfig{file=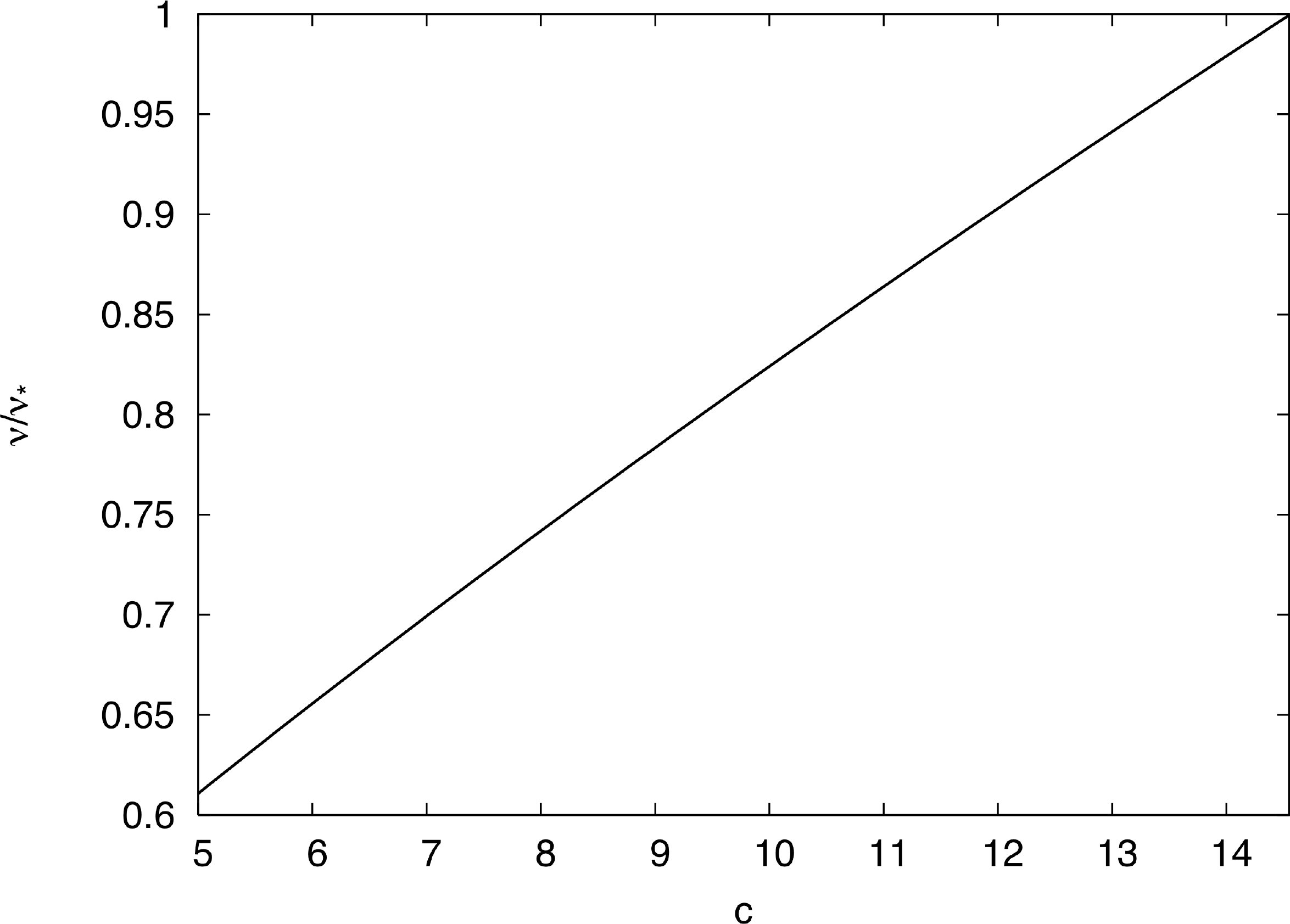, width= 8cm, angle = 0}
\caption{The signal-to-noise with which a NFW halo of concentration $c$ is detected, $\nu$, relative to a halo of concentration $c_{*}=14.55$, 
$\nu_{*}$, as a function of $c$.}
\label{cplot}
\end{figure}

Now to illustrate why differences in concentration would impact on the shear-selected peak counts, we consider the optimal 
signal-to-noise of halos as a function of their concentration, at fixed mass.  In the limit of a filter that is exactly matched to the halo profile, 
and neglecting large-scale structure noise (so concentrating on the most significant cluster mass peaks), the signal-to-noise is given by,
e.g. \cite{berge},
\begin{equation}
\nu=\frac{\sqrt{n_{g}}}{\sigma_{\epsilon}}\sqrt{\int{{\rm d}^{2}\theta\,\kappa^{2}(\theta}}\,
\end{equation}
which they show for an NFW profile is 
\begin{equation}
\nu\propto \rho_{s}r^{2}_{s}\sqrt{G(c)}
\end{equation}
for a given lens redshift and source redshift (or redshift distribution), and number density of source galaxies, with
\begin{equation}
G(c)\approx\frac{0.131}{c^2}-\frac{0.375}{c}+0.388-5\times10^{-4}c-2.8\times10^{-7}c^{2}\,.
\end{equation}
In a given cosmology (fixed critical density, and overdensity required for collapse) since
\begin{equation}
\rho_{s}\propto\frac{c^{3}}{{\rm ln}(1+c)-\frac{c}{1+c}}\,,
\end{equation}
and comparing halos with the same mass (and $r_{200}$), the ratio of the signal-to-noise of halos with different concentrations is
\begin{equation}
\frac{\nu}{\nu_{*}}=\frac{\frac{c}{{\rm ln}(1+c)-\frac{c}{1+c}}\sqrt{G(c)}}{\frac{c_{*}}{{\rm ln}(1+c_{*})-\frac{c_{*}}{1+c_{*}}}\sqrt{G(c_{*})}}\,,
\end{equation}
which gives the signal-to-noise of a halo, $\nu$, compared with a fiducial halo, $\nu_{*}$. In Fig.\,\ref{cplot} this is plotted with respect to the concentration of the Oguri et al. $10^{14}h^{-1}M_{\odot}$ halo; note that the Duffy et al. halo of the same mass would yield 
$\approx 65\%$ of the signal-to-noise due to lower concentration.

With what accuracy can the parameters of the mass-concentration relation potentially be constrained using the peak counts, for a DES-like survey? Let us assume that the Duffy et al. relation is the true, fiducial, description. Considering sources to be at $z=1$, we estimate the range of parameters that would be consistent with the "observed" peak counts, by running different grids of parameters and considering the most significant peaks with $\nu>4.5$. To approximately account for cosmic variance, the error bars in Fig.\,5 are doubled, although the statistics of highly significant peaks is dominated by shot noise (Hu \& Kravstov (2003)). From Fig.\,5, it is already clear that the halo counts from the fiducial model would be inconsistent with an Oguri et al. $M-c$ relation. Given these observational error bars, and keeping the mass and redshift dependence of the $M-c$ relation fixed, the concentration normalisation can be determined to within $\sim 7\%$. Fixing the normalisation and redshift dependence, the data would be inconsistent with a flat dependence on halo mass, but would accommodate a relation that is $\approx 50\%$ flatter or steeper in $M_{\rm vir}$. Finally, for the evolution with redshift, fixing the normalisation and mass dependence, the observed counts are inconsistent with zero evolution of the $M-c$ relation with redshift, but consistent with an evolution that is $\approx 30\%$ flatter or steeper in $(1+z)$. 

In practice one would explore which method of peak detection was best suited to the data set, either by application to simulations, or by testing the recovery of known lower mass clusters. Peak counts looks promising as a technique that complements the determination of the $M-c$ relation from detailed studies of individual clusters (e.g. Comerford \& Natarajan (2007); Oguri et al. (2009); Okabe et al. (2010)), or from statistical studies (e.g. Mandelbaum et al. 2008), in particular when the uncertainty on cosmological parameters would be further reduced by Planck and supporting data. 

\subsection{The recent extension to the halo model by Giocoli et al.}
Recently, \cite{Gio10} extended the halo model formalism to include realistic substructure within individual haloes, and to include scatter in halo concentration at fixed mass. 
With their prescription for substructure, as motivated by the findings of N-body simulations (e.g. \cite{Gao04}), the predicted power on very small scales is increased, bringing the matter power spectrum derived from the halo model formalism into better agreement with cosmological N-body simulations. 

The increase in the contrast between true halo peaks and large-scale structure from this extension would likely lead to an increase in halo counts compared with standard implementations of the halo model.
The extension to the halo model formalism of Giocoli et al. (2010), may be especially pertinent to properly accounting for the weak lensing signal due to low-mass haloes, as they note, and predicting the counts of low-mass haloes.  For our comparison of the difference between peak counts for the Duffy et al. and Oguri et al. $M-c$ relations, any linear structure would be the same in both cases, and the higher normalisation $M-c$ relation yields more peak detections. This could then be further enhanced by the inclusion of substructure. For the detection of the highest significance peaks, where the weak lensing signal is dominated by massive cluster halos, the contrast against the large-scale structure is less of an issue. In addition, simulations show that substructure itself changes the weak lensing signal of these massive haloes by only a few percent (\cite{kss}). Assessing the impact of this extended halo model formalism on weak lensing peak counts is beyond the scope of this paper, and will be the subject of future work. At that point, estimating the accuracy with which the parameters of the mass-concentration relationship can be estimated could be undertaken with a Fisher matrix analysis, also incorporating the (photometric) redshift distribution of sources from which the weak lensing signal is being measured, and combining information from the measurement of the cosmic shear power-spectrum for example.

As discussed further by \cite{gio10b} their extended halo model with the inclusion of  substructure is of much importance in predicting the flexion signal of haloes, and in the investigation of strong-lensing properties such as universal magnification invariants. \cite{gio10b} also note that their formalism could be further refined, for example to include the impact of nonlinear bias and to include halo shapes.

\subsection{Non-Gaussianity}
Some degree of non-Gaussianity in the primordial density field can arise in even the simplest inflationary models as discussed in e.g. \cite{lidlyt}, and we now briefly compare this with the impact of $\sigma_{8}$. Commonly non-Gaussianity is quantified in terms of the parameter $f_{\rm NL}$ such that the Bardeen potential $\Phi$ can be expressed as the sum of a Gaussian random field $\Phi_{\rm L}$ and a quadratic departure term: $\Phi=\Phi_{\rm L}+f_{\rm NL}[\Phi_{\rm L}^2-\left<\Phi_{\rm L}^2\right>]$. The tightest constraints currently come from CMB data - using WMAP 7-year data, \cite{wmap7} find $13<f_{\rm NL}< 97$. \cite{grossi07} carried out dark matter simulations with various values of $f_{\rm NL}$, and found that only modest departures from the halo mass function arise even for large positive or negative values of $f_{\rm NL}$. In the context of weak lensing statistics, and using the same set of simulations, \cite{pace} show that non-Gaussianity also has a minimal impact: for values of $f_{\rm NL}$ within the current bounds, the impact on the cosmic shear power spectrum is well within the uncertainty due to the current bounds on $\sigma_{8}$. \cite{pace} also remark that since non-Gaussianity is comparatively much more effective at small scales, the scale-dependence of observables can be used to aid parameter constraint in future surveys. Indeed, as noted earlier, it has recently been shown that the covariance of cluster counts is very sensitive to primordial non-Gaussianity (\cite{ogurifnl};  \cite{cunha}). Although beyond the scope of this paper, it would be interesting to examine the extent to which the uncertainty on the $M-c$ relation could be reduced by self-calibration of shear-selected peak counts (similar to the scheme of Zentner et al. (2008)) binned in redshift, and allowing for non-Gaussianity at a level consistent with forthcoming Planck data.

\subsection{Baryonic physics}
The impact of baryons on dark matter halos as seen in simulations, compared with the change in profiles due to changing the halo concentration to be consistent with $M-c$ relationships at the same mass, was considered in Section 5. In Mead et al. (2010) we demonstrated that simulated clusters with AGN feedback are less efficient strong lenses (specifically in their production of giant arcs) than clusters simulated without AGN feedback, and are in fact similar to dark matter only halos. Here we considered the $\sim 3\times 10^{14}\,{\rm M}_{\odot}$ cluster from the simulations of \cite{puchwein08}, also used in Mead et al. (2010) where the stellar mass fraction is in very good agreement with observations, in the case where AGN feedback is accounted for alongside other baryonic processes. As shown in Fig.\,\ref{baryoys}, the corresponding FFTs of these density profiles that enter into the halo model predictions illustrate that the simulated model with AGN feedback has much less power on small scales than either the simulated model without AGN feedback (i.e. where there is over-cooling), or of the analytic model with a concentration at the mean of the Oguri et al. $M-c$ relationship. From this illustration, and from the general findings of Mead et al. (2010) and Puchwein et al. (2010) we emphasize that for the observational measurements of the $M-c$ relation made on cluster scales, or correspondingly for individual clusters with very high concentrations, baryons are unlikely to be a significant culprit in yielding results that greatly differ from the expectations of $\Lambda$CDM. Note that the number of simulated clusters, groups and galaxies currently available is too small to specify a global $y(k,M)$ form, but an interim possibility that we defer to later work would be to calibrate an analytic form, such as that used in \cite{guill10}, using a fairly large set of high resolution simulations that reproduce the density profiles and stellar mass fractions from observations.  

\subsection{Conclusions}
Understanding the physical processes, or the observational selection effects, that give rise to some clusters having seemingly high concentrations, and to tension in $M-c$ relations derived from various types of observations, will be possible in the near future. In particular, we have shown that the counts of halos obtained using weak lensing observations, and the cosmic shear power spectrum are sensitive to the $M-c$ relation. Alongside the exciting preparations for observational surveys such as DES, much progress is being made in developing tools that are very useful in better understanding the implications of observations. For example, during the work for this paper, Giocoli et al. (2010) presented an extension to the halo model formalism that incorporates substructure in halos, as well as allowing for scatter in halo concentration at fixed mass.

It will be very interesting to compare constraints on mass profiles and their concentrations from measurements of both weak and strong lensing by individual selected clusters from upcoming surveys, from carrying out statistical studies in DES such as that in SDSS of Mandelbaum et al. (2008), to explore using tomographic cosmic shear with self-calibration (Zentner et al. 2008) and from measurements of weak lensing peak counts. 

\vspace{0.5cm}
\noindent{\bf{ACKNOWLEDGMENTS}}\\
LJK thanks the Royal Society for a University Research Fellowship, and JMGM thanks STFC for a postgraduate studentship award.
We thank Carlo Giocoli, Debora Sijacki, Ian McCarthy, Antony Lewis, Zuhui Fan and Sirichai Chongchitnan for helpful discussions. We thank Ewald Puchwein, Debora Sijacki and Volker Springel for use of their cluster simulations, and the referee for a constructive and helpful report.
\mmm

\bibliographystyle{mn2e}
\bibliography{lensing}

\end{document}